# Collective Dynamics of Water in the Living Cell and in Bulk Liquid. New Physical Models and Biological Inferences[†]

*Eugen A. Preoteasa* [1,¶] *and Marian V. Apostol* [2,‡]

[1] Department of Life and Environment Physics and [2] Department of Theoretical Physics,
Horia Hulubei National Institute for Physics and Nuclear Engineering,
407 Atomistilor, P.O. Box MG-6, 077125 Bucharest-Magurele, Romania

## Abstract

In the frame of collective dynamics in water, models built on elementary excitations and long-range electromagnetic interactions in the cell and bulk liquid are presented.

Making use of the low effective mass of water coherence domains (CDs), we examined the relevance of simple quantum models to cellular characteristics. A hypothesis of CDs Bose-type condensation, and models of CD in spherical wells with impenetrable and semipenetrable walls, and of an isotropic oscillator consisting of two interacting CDs were investigated. Estimated cellular volumes matched to medium-sized bacteria and small prokaryotes, and to some organelles in eukaryotic cells. Also, the cytotoxic effects of heavy water in eukaryotes were explained.

In another approach we proposed a plasmon-like model of a $H^{+z} – O^{-2z}$ ionic stable plasma ($z \sim 0.03$) in liquid water. In addition to plasmonic oscillations, the model predicted sound-like non-equilibrium elementary excitations that we called densitons (the sound anomaly of water), the vaporization heat and the permittivity dispersion, in agreement to experimental data. The fields generated by ionic plasma oscillations in living cells may support long distance intercellular correlations.

---

[¶] Correspondence to: Eugen A. Preoteasa, Horia Hulubei National Institute for Physics and Nuclear Engineering, Department of Environmental and Life Physics, 407 Atomistilor, P.O. Box MG-6, 077125 Bucharest-Magurele, Romania. E-mail address: eugen_preoteasa@yahoo.com, eap@nipne.ro . Tel. +40-21-404 6223, Fax +40-21-432 1701.
[2]
[‡] Marian V. Apostol, E-mail address: apoma@theory.nipne.ro

# Introduction and conceptual background

A fundamentally new approach for understanding the characteristics of the cell and higher organisms based on their integrative features, such as the emergence of self-organization and hierarchies of collective order, has been set up by Fröhlich's model of long-range coherence in living systems (see Fröhlich 1968a,b, 1975, 1978, 1980, 1982). This is consistent to the quantum electrodynamics (QED) model of condensed matter in general and of coherence domains in water as elaborated by Preparata, Del Giudice and colleagues, which explain many remarkable phenomena in the living cell (e.g. Del Giudice 1993, 2008, Del Giudice et al. 1982, 1983, 1986,1988, 2002, Preparata 1995). We suggested previously some possible applications of these concepts to processes occurring in membrane systems based on the involvement of water long-range correlations in cells (Preoteasa and Apostol 2006). Along the above line which aims to account for the collective dynamics in liquid and intracellular water and to look for a deeper insight of the living systems, here we proposed two further developments. We examined the employment of simple quantum models which exploit certain properties of the water coherence domains in order to explain the dimensions of living cells and the cytotoxicity of heavy water, and we suggested a new model of liquid water dynamics based on a two-species ionic stable plasma as a view able to describe several properties of water and potentially relevant for intercellular communication. The conceptual background is outlined in the introductory section.

## *The cell dimensions problem. A simple unexplained biological fact*

Life is strikingly different of the non-living matter, and its characteristics are expressed in the essence at the level of the cell, its basic unit. The most elementary characteristic of cells concerns their dimensions. With a few exceptions, cells are objects of typically 1-100 μm diameter and, conceding also for their inner structures one to several orders of magnitude smaller (Alberts et al. 1994), they place themselves between the microscopic, quantum world and the macroscopic, classical one. The dimension of cells is basically empirical fact from the standpoint of traditional biology. Currently, the molecular biology of the cell postulates a lower size limit by assuming that it is due to a minimum number of different types of enzymes (from several hundreds to one thousand) necessary to perform the vital functions, while the upper size limit is assigned to a surface-to-volume ratio able to ensure a proper ratio between the influx of nutrients and the efflux of catabolites at a given level of metabolism, which in its turn decreases with the size of the cell (Zarnea 1983). This explanation, although ingeniously connecting some distinct cellular characteristics and processes, has to rely on several biochemical data which, however, are essentially of an empirical nature. It rather displaces the problem at a deeper level instead of fully solving it. Therefore the size of cells still seeks for a rationalization in a wider context, and an examination from a more physical perspective appears as an attractive alternative.

A possible answer was given first by Schrödinger (1944), who pointed out that in a living organism molecules must cooperate, and that this condition requires a volume large enough to ensure the cooperation of a sufficient number of molecules against thermal agitation. Along this line the dissipative structures theory postulates a generalized structure-function-*fluctuation* relationship and, looking up the cell as a giant density fluctuation, shows that its dimensions must exceed the Brownian diffusion which takes place during the lifetime of the cell (Prigogine 1980). Such an approach - although it would require the cell life expectation as an input and in spite of its phenomenological character - succeeds to look at the cell as a whole, without taking into account the molecular details inside the cell. And by this it is fundamentally different of the molecular biology point of view, which always gives emphasis for specific details. Indeed, the fact that molecular biology did not provide until now a comprehensive explanation for this straight feature of the cell may be not purely accidental.

*The heavy water cytotoxicity problem*

While full substitution with stable heavier isotopes (such as carbon-13, nitrogen-15, and oxygen-18) in living organisms is generally well tolerated, replacement of $H_2O$ with $D_2O$ has strong effects which, depending on the concentration, may be very severe, especially in eukaryots (Wikipedia – Heavy water, 2008). Heavy water increases the period of circadian oscillations in unicellular organisms, green plants, isopods, insects, birds, and mammals. Heavy water stops eukaryotic cell division. Therefore plants stop growing and seeds do not germinate when given only $D_2O$. Mammals develop (sometimes irreversible) sterility when their body water reaches 25% deuteration, and they die after a week when deuteration approaches about 50%; also high concentrations of heavy water (90%) rapidly kills fishes, tadpoles, flatworms, and drosophila. Because $D_2O$'s general action inhibits the eukaryotic cell division, the symptoms and mode of death in acute deuterium poisoning appear to be the same as that in cytotoxic poisoning (such as chemotherapy) or in acute radiation syndrome. Also, $D_2O$ is more toxic to malignant than normal animal cells. As in chemotherapy, deuterium-poisoned mammals die of a failure of bone marrow (bleeding and infection) and intestinal-barrier functions (diarrhea and fluid loss).

On the contrary, prokaryotic organisms such as bacteria (which do not have the mitotic problems induced by deuterium) may be grown and propagated in fully deuterated conditions, resulting in replacement of all hydrogen atoms in the bacterial proteins and DNA with the deuterium isotope (Kushner et al 1999).

Within the frame of molecular biology the mechanisms of $D_2O$ cytotoxicity are largely unexplained (Pittendrigh et al 1973). However we note that the differences of $D_2O$ effects in eukaryots and prokaryots correlate to the fact that the former cells are generally larger and have compartmentalized internal structure. From this point of view we meet again the question of cell dimensions, and we aim to search for a related approach in order to understand better the problem of heavy water cytotoxicity.

*Some unique features of life*

In contrast to its practical achievements, many unique features of life escape molecular biology. The cell molecules perform a collective dynamics, based on the interaction of many degrees of freedom; otherwise, fluctuations would prevail progressively and destroy the cell before it can multiply and perpetuate itself. There is a general agreement that life is a metastable state. Moreover cells are a special kind of heterogeneous soft matter, displaying basically different types of order, local and global, disposed in a hierarchy on successive scales from nanometer-sized elements to the size of the cell itself and beyond, and involving short- and long-range correlations altogether. At the same time living organisms are open and dissipative structures, optimal, self-regulating, homeostatic systems which work by non-linear and cooperative interactions, and their dynamics is highly integrated.

*On the limits of mainstream molecular biology*

The efforts of the prevailing molecular biology (e.g. Monod 1970, Alberts et al. 1994) to understand the above features of life fail to a large extent, because of its attempt to reduce cell dynamics exclusively to the movement of molecules looked at as classical particle-like objects, isolated excepting for short-range interactions, and obeying always Maxwell-Bolzmann statistics. Here are some principle limits of the reductionist view of molecular biology. 1) *Methodologically controversial.* This is obviously so due to the claim to explain completely the whole by its parts. 2) *Conceptually hybrid and incomplete.* The molecular structure of the cell taken as the *dynamical* initial conditions of the system, while looked at as the expression of *biological* evolution-codified information. As the first protocell in its native form cannot be traced on Earth, the evolution pathway to the actual organisms could never be explained in full detail and without appreciable incertitude, and thus a dynamical description is necessarily incomplete. 3) *Non-decidable in its ultimate predictions.* By postulating the genesis of life on Earth to be dependent of an unique and pure accidental set of conditions, Monod (1970) paved the way towards *'Biological Numbers'*, practically out of the reach of experimental check. Thus in $5.10^9$ years the probability of random synthesis of the about $10^3$ enzymes of the simplest cell was evaluated to $\sim 10^{-40,000}$ (Hoyle and Wickramasinghe 1981); and the probability of $\sim 10^{-24,000,000}$ for the advent of man (Barrow and Tipler 1986) may seem an optimistic estimate, while a more pessimistic and older view gives $\sim 10^{-1011}$ even for the probability of the birth of a simple bacterium like *E. coli* (Morowitz 1968)! However, by contemplating such numbers which are implicit in the Monod-type approach, Jacob, another eminent molecular biologist and a Nobel Prize colleague of the former, admits that "pure chance is not enough" (Thaxton et al 1984) and that the uniquely rich phenomenology of life requires some integrative principle beyond the mere sum of parts and random game of the individual constituents (Jacob 1970).

*Fröhlich's model of long-range coherence in living systems*

A most plausible answer to problems such as the origin of unicellular organisms and man and accounting for the key features of life has been given by Herbert Fröhlich's in his kinetic rate model of long-range coherence in living systems (Fröhlich 1968a,b, 1975, 1978, 1980, 1982). This theory is consistent to the dissipative structure theory but essentially open to a microscopic approach. Fröhlich postulated that non-linear interaction between longitudinal elastic modes / acoustic (phonon) vibrations and electric polarization fields when pumped by metabolic (chemical) energy above a critical threshold results in excitation of coherent electric polar oscillations; non-linearity is involved also in the coupling of the system to a heath bath, which adds thermal noise due to physiologic temperature. The resulting *single coherently excited polar mode* is a special dynamical order based on correlations in momentum space which describes the living *vs.* non-living basic difference; consequently, the living system reaches a *metastable* minimum of energy. The single collective dynamic mode of electric dipoles is excited in the $10^{11}$-$10^{12}$ Hz band and generates intense electric fields allowing long-range Coulomb interactions. Moreover, it "has considerable similarity with the low-temperature condensation of a Bose gas" (Fröhlich 1968b); the system reaches a state that can be termed a Bose–Einstein-like condensation, not in a phase in equilibrium, but in a nonequilibrium dissipative structure (Del Giudice et al 1982). The Fröhlich kinetic rate equation has been later derived on a microscopic basis, for a system of oscillating electric dipoles described by a Hamiltonian written with cell, heat-bath, and energy-pump creation and annihilation boson-type operators; this treatment indicated also that Bose-type condensation may occur in biosystems in a frequency domain under certain conditions (Wu and Austin 1977, 1978a,b). The phonon condensation in the Fröhlich model has been questioned (Yushina 1982), but it has been shown that the non-thermal excitations can still exist in a regime in which the Bose condensation fails (Mills 1982, 1983), and that the elastic forces stabilise the polar modes and lower the threshold for Bose condensation, while the metastable state shows ferroelectric properties on increasing the pumping rate (Bhaumik et al 1976, 1977). In addition to providing a key for the characteristic type of order in living cells, the Fröhlich model succeeded in understanding some overall biological phenomena, such as self-preservation by metabolism, cell-cell signaling, contact inhibition, cancer proliferation, brain waves, long-range interactions between cells and biopolymers, and enzymatic reactions (Tuszynski 2004). Various experimental data sustain the theory, including for instance non-thermal effects in *E.coli* found by Raman scattering (Webb et al 1977, Webb 1980) – thoroughly analyzed by the Milano group (e.g., Del Giudice et al 1982) and reviewed by Pokorny et al (1991) –, increased growth of yeast cells by millimiter waves irradiation at specific frequencies (Grundler and Keilman 1983, 1989), dielectrophoresis of insulating particles by living cells (Pohl 1983), and rouleaux formation of human erythrocytes (Rowlands et al 1981, Rowlands 1983, Paul et al 1983, Fritz 1984). Finally we mention that the genesis of life becomes a much more probable and plausible phenomenon within the Fröhlich model. In systems where clustering occurs, even so limited such as dimer formation, the non-equilibrium stationary state of pumped phonon ensemble is sufficient to give rise to *limit-cycle* behavior (Poland 1989); and this state has the properties of the *terminal state* of the system for all initial conditions (Duffield 1985). (We note that this view supports Gould's concept of „God playing loaded dice" for biological self-organization – see e.g. Gould 2002.)

*Biological order and water in the living cellular and in pure bulk liquid.*

Cells - even the most simple - show an unparalleled complexity, understood as dynamical order (almost) without repetition. The spatial non-repeating *structure* of proteins and nucleic acids and the short-range specific interactions of molecular biology *per se* may not explain the concerted macroscopic collective *dynamics* of molecules in the cell. A vehicle for long-range specific interactions, which might *convert position-space correlations to momentum-space correlations*, appears to be necessary. One can suppose that the all-pervasive cellular constituent – water – may provide the non-repeating information-carrying biopolymers with the environmental support necessary for the emergence of integrated dynamical macroscopic complexity of the living cell. Such a role, however, would require a capacity for water molecules, although having the same chemical structure, to organize in heterogeneous, differentiated dynamical arrangements. In fact overwhelming evidence demonstrates a diversity of altered physical properties of water near the surfaces of biopolymers and cell membranes, as well as heterogeneities in pure bulk liquid water which explain its well-known physical anomalies (Pauling 1982, Chaplin 2008). As it is well-known, the latter make life possible. At the same time, the physical anomlies of pure bulk liquid water encouraged Röntgen to postulate the phenomenological two-phase model of water – in which flickering "ice-like" clusters are surrounded by a normal fluid (Röntgen 1892). Many years latter, this concept and its consequences in the cell were analyzed in the framework of a microscopic theory consistent to the Fröhlich model, briefly presented below.

*The coherent domains QED model of water*

A quantum electrodynamics (QED) theory to account for the properties of condensed matter in general was developped by Preparata (1995). In the QED picture (Del Giudice 1993) both matter and radiation are regarded as "quantum fields", namely as objects fluctuating in space and time and showing the wave-particle complementarity. In the QED approach of water, an esential prerequisite is that $H_2O$ molecules possess continually fluctuating dipole moments and that fluctuating dipole moments radiate electromagnetic fields. In a two-body interaction between atoms or molecules mediated by electromagnetic (EM) field, only static fields are important. But in the general case of N-body interactions, the time-dependent radiative part of the EM field is no longer negligible and, as a result, the small individual fluctuations of many individual components may superpose coherently (in phase). Considering an ensemble of N molecules mutually coupled through an EM field, the fluctuations of the EM field whose frequency matches the energy difference between two energy levels of the molecules can couple to corresponding fluctuations of the matter field. Beyond a threshold of molecular density, the coupled fluctuations of both the EM and matter fields will be damped no longer, and the two fields will share a common oscillation. Thus a coherent oscillation of matter and EM fields arises, described by a unique wave function in which it is impossible to trace the individual components.

To understand the phenomenological two-phase model of water, the QED approach (Del Giudice et al 1988, Preparata 1995) assumes that the cluster-forming $H_2O$ molecules are in a ground state that differs from the ground state of the isolated molecule. In this new ground state, the $H_2O$ electron cloud is deformed, the effective radius and the dipole moment are increased, so as to generate the hydrogen bonds which stabilize the clusters In the clusters *the water dipoles oscillate coherently, in-phase*; the clusters are *coherent domains (CD)*. Thus pure bulk liquid water consists in two interspersed phases, *coherent* and *incoherent*. The incoherent phase comprises water molecules in the molecular ground state (as observed in the gas phase) packed in a highly dense state in the interstices around large clusters in which the water molecules perform hindered rotations and interact coherently with a large electromagnetic field. (The shape and dimensions of the CD may apparently vary from filaments of 75 nm diameter to spheres of 10 μm diameter.) Thus the coherent phase is built up by the collection of such "coherent domains" that at a given temperature have survived the disordering attacks of thermal fluctuations. In the coherent phase, due to the oscillations of the water molecules between the ground state and an excited state at 12.06 eV – whose relative probabilities are about 0.87 and 0.13 – the volume occupied by each molecule is definitely larger than the volume occupied by the molecules of the incoherent phase, thus rendering the coherent phase density (0.92 g $cm^{-3}$) much closer to the density of ice. The energy gap that protects the molecules from "evaporating" from the coherent phase into the incoherent, "gaseous" one, allows to assume a roton-like excitation phase of "liquid" water. Also, a difference in the electrostatic energy between coherent and incoherent phase is expected and evaluated to a small negative difference $\delta_{es} \sim -0.022$ eV (Preparata 1995).

Del Giudice and colleagues (1986) extended the QED model of water to living cells. In the cell, the water dipoles interact with the polar surface of biopolymers, and their symmetry, spherically invariant O(3) when free in isotropic space, breaks down to the U(1) subgroup of axial rotations when a privileged direction appears, leading to collective Goldstone modes of polarization, namely water coherence domains. These are *bosons* (integer spin), with a *low effective mass* (excitation energy) of ~13,6 eV. The coherence domains are shaped as filaments (of 29 nm diameter) and are located around chain-like proteins. At the surface of water CD filaments strong electric field gradients appear, developing *frequency-dependent, specific, non-linear forces* to dipolar biomolecules, of the form (Askar'yan 1962, Del Giudice et al 1983, 1986):

$$F \sim \frac{\omega_0^2 - \omega^2}{(\omega_0^2 - \omega^2) - \Gamma^2} \cdot \nabla E^2 \tag{1}$$

Here $\omega_o$ is the oscillation frequency of biomolecule, $\omega$ – the coherent oscillation frequency in the water filament, $\Gamma$ – the damping of $\omega_o$ oscillations, E – the electric field around the water coherence domain. These forces can bring non-diffusively into contact specific biomolecules, and thus co-ordinate the cell metabolism.

The above long-range forces may provide an explanation also for rouleaux formation of erythrocytes (Rowlands et al 1983) and dielectrophoresis of small dielectric particles by cells (Pohl 1983). Also, long-range interactions may be invoked to explain the recently observed red shift and half-peak bandwidth variation of the optical spectra of a nutrient in suspension of yeast cells as a function of cell density (Bercu et al 2006). Moreover, the water CD model was proposed (Del Giudice et al 2002) as one of the theoretical explanations of a large body of experimental data involving the ion cyclotron resonance (ICR) phenomenon (Blackman et al 1985, Liboff et al 1987; Liboff 1997). Such are the Zhadin experiment, *i.e.* sharp changes in conductivity for amino acids in solution exposed to ICR magnetic fields tuned to *q/m* of amino acid (Zhadin et al 1998), and its extension showing that magnetic fields tuned to amino acid ICR can hydrolyze proteins (Novikov and Fesenko, 2000). The CD interpretation supposes that ions trapped around coherent domains are released when energy of appropriate frequency is applied and then removed by Lorentz forces, by a temperature-sensitive process. In the case of pure water, the same hypothesis supposes that hydrogen ions are extruded to the surface of CDs, and may explain the results of studies in progress evidencing pH changes of water in electric fields applied from outside of the liquid container (Del Giudice 2008).

## Simple quantum mechanical models of coherent domain dynamics in the cell

We retain two properties of the QED model of water coherence domains – boson nature and low effective mass (about 12.1-13.6 eV) – to suggest simple quantum models aiming to explain some characteristics of the cell such as its size and the cytotoxicity of heavy water. In the following we choose the value of 13.6 eV for the effective mass of a coherence domain and approximate a cell (or a mitochondrion) to a spherical cavity without an internal structure and filled with water instead of cytoplasm, by analogy to an unilamelar liposome, and neglect the molecular structure of the membrane.

### *The hypothesis of free water coherent domains' Bose condensation: The minimum volume of the cell*

In search for an explanation of certain basic attributes of life which may involve coherent dynamics and escape molecular biology, one can speculate that such characteristics might be the expression of very unconventional properties, such as imagined for the cell if looked at as a special type of a macroscopic quantum object. In such systems, the particles involved in their dynamics are bosons, and their coherent states expressing quantum properties at macroscopic level emerge by Bose-Einstein condensation. This kind of transition occurs only below a critical temperature $T_c$ where the individual wavefunctions collapse together in an unique wavefunction with a single phase, as the momentum of the particles falls down and their de Broglie wavelength increases above the average distance between them. The Bose condensation takes place in superfluid liquid helium and superconductors only at very low temperatures, but this is due to the relatively high mass of the $^4$He atoms and of the Cooper electron pairs while the density has values close to normal in liquids and solids. In the core of stars, it may arise at very high temperatures, but this is owed to a very high density of matter. However in lasers, while pumped with energy from outside, condensation of photons from the resonant cavity occurs at normal densities and at room temperature, due to the sufficiently low effective mass of photons.

The effective mass $m_{eff}$ of the coherent polarization domains of water in cells, taken of only $\cong 13.6$ eV $\cong 2,4242 \times 10^{-35}$ kg, is appreciably lower as compared *e.g.* to the electron mass (511,000 eV); accordingly, at a given translation speed, the CDs will have a much longer de Broglie wavelength $l = h/m_{eff}v$ and enhanced wavelike properties. On this ground a hypothesis related to the Bose-like condensation in cells can be examined. The cell may be looked up as a spherical cavity of volume $V$ containing a 'gas' formed of $N$ 'free' water CDs moving inside, whose wavefunctions reflect totally on the membrane. Here we completely ignore the specific structure and properties of the cell membrane. Due to the low mass, the critical temperature $T_c$ for the condensation of the above CD boson gas may exceed the ordinary temperature of biological organisms ($\cong 37$ ºC $\cong 310$ K), if reasonable values of the CD density is assumed. Then, in agreement to the "metabolism first" hypothesis of Dyson (2000) we infer that the living state of a cell is essentially defined by its synchronized metabolic reactions (and not by its multiplication). We further suppose that this ensemble of biochemical reactions, in order to be dynamically concerted and synchronized as they are, requires non-diffusive co-ordination by the action of the frequency-dependent, specific, non-linear forces between water CDs and dipolar biomolecules, as given by eq. (1) and postulated by Del Giudice. In this way the water CDs will provide in the cell a certain type of Maxwell's demons; recently the possibility of quantum Maxwell's demons has been examined (Kieu 2004). In the protocell from the first moments of evolution, this role will be crucial as an alternative for specific enzymes before the advent of the latter. Moreover we assume that the transition from chemical dynamics in a vesicle *to the living state* of the cell at its basic degree_occurs when the co-ordination of biochemical reactions reaches a point above a threshold, and that this critical level of dynamical order in the cell is reached when a Bose-type condensation of the free water CDs in the cytoplasm arises. Thus life, manifesting itself by new macroscopic dynamic properties, may be viewed as a coherent state of CDs emerging in the whole volume of cell. And with this image a rough estimation of the minimum cell volume is possible.

To reach this end, we employ the Bose-Einstein equation (Bose 1924; Einstein 1925, Landau 1941, Landau and Lifchitz 1980), which gives the critical temperature $T_c$ for the condensation into the lowest accessible (ground) quantum state of a uniform three-dimensional gas consisting of non-interacting bosons with no apparent internal degrees of freedom gas:

$$T_c = \zeta(3/2)^{-2/3} \left(\frac{N}{V}\right)^{2/3} \frac{h^2}{2\pi m_{eff} k_B} \qquad (2)$$

Here in our case $N$ is the number of free CD bosons in the cell, $V$ is the volume of the cell, and $m_{eff} \cong 13.6$ eV $\cong 2.4 \times 10^{-35}$ kg is the effective mass of the water CD; $h \cong 6.626 \times 10^{-34}$ J·s $\cong 4.136 \times 10^{-15}$ eV·s is Planck's constant, $k_B \cong 1.380 \ 10^{-23}$ J·K$^{-1}$ $\cong 8.617 \ 10^{-5}$ eV·K$^{-1}$ is the Boltzmann constant, and $\zeta(3/2) = 2.6124$ is the Riemann zeta function. In general the condition of non-interacting bosons is not true (e.g., Van der Waals interactions exist between $^4$He atoms), and an accurate description of the state of the Bose-Einstein condensate is in terms of its wavefunction written as a solution of the Gross-Pitaevski equation (Pitaevskii and Stringari, 2003). This is a non-linear Schrödinger equation postulated by treating the bosons within a mean field theory, and which takes into account the inter-particle interactions as well as an external potential. Due to the long-range Coulomb interactions, the free CD bosons in the cell hardly could be non-interacting. Also, the 'internal' degree of freedom associated to the coherent dipole oscillation is neglected. However, eq. (2) provides a suitable approximation to evaluate the cell volume,

$$V \geq V_{min} = \frac{N}{\varsigma(3/2)} \cdot \frac{h^3}{\left(2\pi \, m_{eff} \, k_B \, T_C\right)^{3/2}} \tag{3}$$

To find $V_{min}$, we assume that the Bose-type condensation occurs in the cell for $N = 2$, which is the minimum possible number of water CDs for a meaningful condensation. Thus for $T_c = 310$ K, *the minimal volume of a living cell* is estimated to:

$$V_{min} = 0.42 \ \mu m^3 \tag{4}$$

and at this density the volume required per free water CD in the cell is obviously 0.21 $\mu m^3$.

The estimated minimal cell volume is definitely higher than that of the smallest known cell, the bacterium *Mycoplasma,* although the volume of the latter seem to vary within large limits (from 0.004 to 0.35 $\mu m^3$ depending of its mean radius, assumed between 0.06 to 0.44 $\mu m$). The same holds true for the spherical blue-green alga *Prochlorococcus* of the *Cyanobacteria* genre, of $0.1 - 0.3 \ \mu m^3$ volume. However another typical prokaryotic cell, *E. coli*, with a volume of $0.39 - 1.57 \ \mu m^3$ estimated for a radius of $0.5 - 1 \ \mu m$ and a lenth of 2 $\mu m$, approximately coincides with the calculated minimal cell volume; note that, supposedly, $N = 2 - 7$ for these volumes. Moreover the condition V > $V_{min}$ is entirely fulfilled for some taxonomic groups of bacteria, such as *Eubacteria*, *Myxobacteria* and most *Cyanobacteria*, with volumes of $1 - 5 \ \mu m^3$ or more (Zarnea 1983). Probably, the first protocells were of about the size of such small and typical prokaryotic cell, as this allowed the transition to the coherent state of life for a very small number of free water CDs. The eukaryotic cells with much larger volumes (85 $\mu m^3$ for the erythrocyte, 270 $\mu m^3$ for the lymphocyte, 4200 $\mu m^3$ for the macrophage) easily verify the minimum volume criterion. Thus in similar conditions of temperature and CD density, the eukaryotic cells will require $10^2 - 10^4$ free water CDs. In an erythrocyte about 400 free water coherence domains would be expected; our image favours the view of the red blood cell as a living entity and not as a merely 'functional' system, in spite of the fact that it loses the genetic equipment during erythropoiesis.

From a biological standpoint our hypothesis requires further analysis, as the total number of coherent polarization domains of water in cells may be appreciably higher, if the water domains associated to the chain-like proteins with ATPase activity are included (Clegg 1983) in addition to the free water CDs we considered. However, those many protein-associated water CDs may play a role only in complex cells where the free water CDs may be insufficient to co-ordinate metabolism. From a physical point of view we note that the free CDs in our approach are similar in nature to the coherent domains formed without metabolic energy supply in pure bulk liquid water (Del Giudice et al 1986, 1988, Preparata 1995); however, we assumed the free CDs to occur in the complex and diversified polar environment of the cytoplasm, where water is interfacially organized. Moreover, without the energy pumping, our condensate of water CDs is reached in a phase at equilibrium, and not in the nonequilibrium dissipative structure which characterizes the living state, as pointed out in the Fröhlich model (Fröhlich 1968a,b, 1975, 1978, 1980, 1982). But in spite of all the approximations involved, the pretty accurate prediction of the minimal cellular volume encourages us to believe that the above simple model succeeds to seize some real features of the living cell.

*The hypothesis of water coherent domains in a spherical well: The maximum cell volume*

Taking advantage of the low effective mass ($m_{eff}$ = 13.6 eV = 2.4242 × 10⁻³⁵ kg) of the water coherence domains, a quite different approach allowed us to evaluate also an upper limit for the cell volume. The collective dynamics of the water CDs inside the cell is now neglected in favour of a simple uniparticle model. In addition to coherent oscillations of water dipoles, a CD may have additional freedom degrees, such as translation, rotation and deformation. The cell is described as a spherical potential well of radius $a$ (Gol'dman et al 1960, Landau and Lifshitz 1966, Messiah 1969) within which the translational wavefunction of a water CD reflects on the membrane. To a first approximation, the orbital movement is neglected ($l = 0$). Obviously, here the postulated spherical shape of the cell is a very coarse approximation to the real one, except in the case of a few type of cells, like the cocci bacteria and the yeast.

In the simplest case of the coherence domain inside a spherical well with *impenetrable* walls for (namely, *infinite* potential barrier, $U_0 \rightarrow \infty$), the translation energy of the CD is quantized on an infinite number of discrete levels $E_1, E_2, \ldots$ :

$$E_n = \pi^2 \cdot \frac{\hbar^2}{2\,m_{eff}\,a^2} \cdot n^2 = 9.87 \cdot u \cdot n^2 \qquad (n = 1, 2, ...) \qquad (5)$$

where $\hbar = h/2p = 1.05457 \times 10^{-34}$ J·s = 6.58212 × 10⁻¹⁶ eV·s is the reduced Planck's constant, and we used the notation:

$$u = \hbar^2/2m_{eff}a^2 \qquad (6)$$

(for example, $u = 2.29 \times 10^{-22}$ J = 1.43 × 10⁻³ eV for $a = 1$ μm). If the cell is described as a spherical well with *semipenetrable* walls, an arbitrary but reasonable numerical value should be given to the *finite* potential barrier, e.g.

$$U_0 = 4u = 4\,\frac{\hbar^2}{2\,m_{eff}\,a^2} \qquad (7)$$

The corresponding quantized energy levels are:

$$E_n = 1.155 \cdot \frac{\hbar^2}{2\,m_{eff}\,a^2} \cdot n^2 = 1.16 \cdot u \cdot n^2 \qquad (n = 1, 2, ...) \qquad (8)$$

where the coefficient 1.16 was determined by solving an algebraic transcendent equation well-known from the formally similar case of unidimensional well with finite potential. For $m_{eff}$ = 13.6 eV and a spherical well of 2 μm diameter, $a = 1$ μm, i.e. of the size of a prokaryotic cell, the energy $E_1 = \hbar\omega_1$ of the first level in the two above cases scales as 3.5 × 10¹² Hz for an impenetrable wall and 4.0 × 10¹¹ Hz for the semipenetrable wall. Although the single collective dynamic mode of electric dipoles was not considered explicitly, note that the estimated values agree as order of magnitude to the frequency band of coherent oscillations predicted by Fröhlich (10¹¹-10¹² Hz). The similar energies involved in the translation of a CD inside the spherical well suggest some correlation between the two freedom degrees; and our simple model appears to be relevant at least to some degree, in spite of its rough approximations.

To estimate the maximum volume of a cell, we postulate that the quantum state of the CD in the spherical well should be stable. In this way we get closer to the stability requirements of the collective approach – namely the incidence of the Bose-Einstein condensation of the free water CD three-dimensional 'gas' into the lowest accessible quantum state, and the metastable nature of the living state as postulated by Fröhlich. The stability requires that the second level $E_2$ of the CD in the spherical well should be thermally inaccessible from the first (ground) level $E_1$, or that the energy difference $E_2 - E_1$ should exceed the thermal energy at physiological temperature 37 °C = 310 K. For the cell described by the spherical well [with impenetrable walls] this writes:

$$\alpha^2 \frac{\hbar^2}{2\,m_{eff}\,a^2}\left(2^2 - 1^1\right) \geq \frac{3}{2}\,k_B\,T \tag{9}$$

where $\alpha = \pi = 3.1416$ for the cell with impenetrable walls and $\alpha < \pi$, in particular $\alpha = 1.0747 = 1.155^{1/2}$ for the cell with semipenetrable walls of height $U_0 = 4u$. Hence *the maximum radius* of the cell is evaluated as:

$$a \leq a_{\max} = \frac{\alpha \cdot \hbar}{\sqrt{m_{eff}\,k_B\,T}} \tag{10}$$

Therefore $a \leq 1.02$ μm for impenetrable walls and $a \leq 0.35$ μm for the considered case of penetrable walls. The corresponding *maximum volume* values of the spherical cell ($V_{max} = 4\pi/3\ a_{max}{}^3$) in the two cases are 4.45 μm³ and 0.18 μm³, respectively.

The first maximum value $V_{max} = 4.45$ μm³ was obtained on assumption of impenetrable walls (total reflexion of the wavefunction) just as the minimum volume estimated above by the hypothesis of a Bose-type condensation; taken together, they establish a criterion for the *limits of the cell volume*:

$$0.42\ \mu m^3 = V_{min} < V_{cell} < V_{max} = 4.45\ \mu m^3 \tag{11}$$

This range of values for the volume of a cell is rather well confirmed by the biological data (Zarnea 1983) not only as order of magnitude, but even quantitatively within the variability of individual cells for certain typical prokaryotic cells such as *Eubacteria* and *Myxobacteria* (1 -5 μm³) and *E. coli* (0.39 – 1.57 μm³), and for some smaller species among *Cyanobacteria*. However a more refined model may be necessary for prokaryotes with volumes above 4.45 μm³, such as the gigantic bacterium *Achromatium* (~8000 μm³ ).

The second value $V_{max} = 0.18$ μm³ was found assuming semipenetrable walls, while the minimum volume was estimated by the Bose-type condensation hypothesis assuming total reflexion of the wavefunctions on the cell walls, that is, impenetrable walls; therefore they can not be used together to define volume limits. Nevertheless, the biological data on the smallest prokaryotes, e.g. *Mycoplasma* and *Rickettsia* (0.01 – 0.03 μm³), the blue-green alga *Prochlorococcus* (0.1 – 0.3 μm³), etc., match rather well a maximum volume of 0.18 μm³. A careful analysis is needed to correlate the above differences between the small and medium size bacteria to their morphological and biochemical characteristics.

Apparently, the upper calculated limits for the cell volume are not confirmed for the much larger eukaryotic cells (~$10^2 – 10^4$ μm³; for humans, 300 – 15,000 μm³ as a rule, depending on the tissue; and 8,200,000 μm³ for the human ovocyte, the largest human cell).

However, the eukaryotic cells are highly compartmentalized, and organelles divide the cell interior into small spaces, that by and large meet with the above volume limits. Thus the mitochondria have volumes of the order of $0.8 - 8$ $\mu m^3$ and the cisternae of the Golgi apparatus are about $0.7 - 6$ $\mu m^3$. The quantitative deviations of our values $(0.4 - 4.5$ $\mu m^3)$ from the experimental ones are not very large and they may be easily attributed to the neglect of the real shape of organelles, e.g. ellipsoidal for the mitochondria and flattened, distorted cylinders or disks for the Golgi cisternae. Also, the real situation is complicated by the incidence of quasi-parallel and quasi-periodic membrane structures inside the organelles, such as the cristae in mitochondria and the thilakoids in chloroplasts. Therefore the predictions of our simple models appear surprisingly good. We note finally that the estimated cell volume limits provide further biologically relevant insight, as they point to the similarity of the mitochondria to the prokaryotic cells; this approximate match implicitly sustains the hypothesis of evolutionary internalization of certain organelles as small foreign cells. Moreover the dimensions of the first protocells may have been in a range close to the evaluated limits.

*The hypothesis of water coherent domains in a spherical well: the toxic effect of heavy water*

The applicability of our simple spherical well model to the eukaryotic cells, endorsed by their internal structure which divides the interior space into sufficiently small volumes, provides a possible explanation for the biologically adverse effects of heavy water. Thus although $D_2O$ and $H_2O$ are chemically identical and most of their physical properties differ by less than $\sim 5 - 10$ %, $D_2O$ induces important biological effects, more severe in eukaryotes than in prokariotes; as mentioned before, mice can even die when fed only with $D_2O$ for a longer period. To search for an explanation, we should take into account factors which differ most, both at the level of eukaryotic *vs.* prokaryotic cells, and at the level of heavy vs. normal water molecule. Thus we may assume that the toxic effects of $D_2O$ in eukaryotes might be related to compartmentalization, which is their basic difference to prokariotes. Also, we may suspect that the difference in the momentum of inertia $I$ of the two isotopic forms of water might be involved, because it amounts approximately from simple to double (as D mass is about twice H mass while the bond lengths remain roughly unchanged), in contrast to most other molecular parameters.

$$I(D_2O) \cong 2I(H_2O) \qquad (12)$$

In fact the mass of the deuterium atom is about twice that of hydrogen, $m_D \cong 2m_H$, while the geometry of the $D_2O$ and $H_2O$ molecules (and hence the length $l$ of the normal from the hydrogen atom to the rotation axis passing through the oxygen) is approximately the same, so that $I(D_2O) = \Sigma\ m_D l^2 \cong 2\Sigma\ m_H l^2 =.\ 2I(H_2O)$. This leads to radical differences in the physical characteristics of the coherence domains in $D_2O$ and $H_2O$, as evidenced by a few simple relations established in the QED theory (Del Giudice et al 1986, 1988). They refer to the fundamental rotation frequency $\omega_0$ of the water molecule rotator:

$$\omega_0 = \frac{\hbar}{2I} \qquad (13)$$

to the size $\delta$ of the coherence domain

$$\delta \sim \frac{2\pi}{\omega_0} \qquad (14)$$

and to the effective mass $m_{eff}$ of the coherence domains:

$$m_{eff} = \frac{\hbar}{c\,\delta} \qquad\qquad (15)$$

where $c$ is the speed of light in vacuum ($c = 2.9979 \times 10^8$ m·s$^{-1}$). Using the momentum of inertia relationship (13) we get immediately $\omega_o(D_2O) \cong \omega_o(H_2O)/2$ and $\delta(D_2O) \cong 2\delta(H_2O)$ and, most importantly for our approach,

$$m_{eff}\left(D_2O\right) \cong \frac{m_{eff}\left(H_2O\right)}{2} \qquad\qquad (16)$$

Thus the substitution of $H_2O$ by $D_2O$ will lead to the reduction of the effective mass of the water CD to its half. Therefore deep consequences are to be expected in the translational dynamics of the coherence domains inside the cellular spherical well. To look after them from a non-reductionist standpoint, we assume that the toxicity of $D_2O$ affects the eukaryotic cell as a whole, and not as a conglomerate of isolated and non-communicating smaller compartments. Therefore we will account for the compartmentalized structure of the eukaryotic cell by considering the translational movement of the water CDs inside contiguous spherical wells with *semipenetrable* walls. This allows for the wavefunction to extend as well beyond the walls of the well (where it decreases exponentially with the distance). It leads also to the question on how the water CD "penetrates" the potential barrier of the membrane. We obviously do not suppose the water molecules of the CD to cross the membrane; we assume only that the dynamical order of the CD may pass from one side to another by a quantum effect, like *e.g.* tunneling. However the detailed nature of the later process is beyond the scope of this study and, once admitted its possibility, we do not need to investigate it here. For our model, it is sufficient to assume that a water CD trapped in a closed volume bordered by inner membranes in the eukaryotic cell can be described as a particle inside a spherical well with a *finite* potential barrier. We use the same height of the barrier as in the example discussed earlier, $U_0 = 4$ $\hbar^2/2m_{eff}a^2 = 4\,u = const.$, both for $H_2O$ and $D_2O$. In the spherical well with finite potential barrier there is a minimum depth for which the energy is quantized, namely for which the first level occurs. This is given by:

$$U_{min} = \frac{\pi^2}{4} \cdot \frac{\hbar^2}{2\,m_{eff}\,a^2} = 2.467u \qquad\qquad (17)$$

Here $m_{eff} = m_{eff}(H_2O)$ for $H_2O$, while for $D_2O$, $m_{eff} \cong m_{eff}(H_2O)/2$, so that t*he minimum depth of the well* for which the first level occurs will be *twice larger for $D_2O$* than for $H_2O$. This evidences a fundamental difference in the relation between $U_{min}$ and $U_0$ for the two isotopic forms of water:

$$\begin{aligned} U_{min}\left(H_2O\right) &\approx 2.5\,u < 4\,u = U_0 \\ U_{min}\left(D_2O\right) &\approx 5\,u > 4\,u = U_0 \end{aligned} \qquad\qquad (18)$$

Therefore – in contrast to the CDs of the $H_2O$-embedded eukaryotic cell – the CDs of the $D_2O$-substituted cell in their ground state *will be no longer in a bound state* with a maximum localization probability at the centre of the spherical well; they will migrate more or less freely over the whole cell, and their translation energy will not show the discrete spectrum associated to the small intracellular compartments. This fundamental, *qualitative* difference in the translation dynamics of the $D_2O$ CDs inside the cellular organelles may provide a possible explanation to the heavy water toxicity in eukaryotic cells, if we assume that a discrete spectrum of the coherent domain translational motion is necessary for the living state. We note, however, that the energy levels will be quantized according to eq. (5) if the cytoplasmatic (outer) cell membrane is supposed to be an infinite barrier to the CDs also in the $D_2O$ environment. This assumption, stating finite *vs.* infinite potential barrier in the case of inner *vs.* outer cell membranes, is consistent to biological data of eukaryotic cells, as the inner membranes are significantly more hydrophilic than the outer ones; thus the protein/lipid ratio of the internal mitochondrial membrane is about twice higher than for the plasma membrane. But for a $D_2O$ CD moving inside an eukaryotic cell of say ~10 μm radius the energy levels will be two orders of magnitude lower than for a $H_2O$ CD restricted within a cell organelle of ~1 μm radius, and thus far below the Fröhlich band ($10^{11}$-$10^{12}$ Hz); this dissimilarity might be involved as well in the heavy water toxicity.

*The hypothesis of a two coupled water coherent domains as a harmonic oscillator and the maximum cell volume*

The simple model of a water coherent domain in a spherical well discussed above is confronted with one limitation of principle, namely its uniparticle nature, which is at variance to the previous many-particle model of Bose-type condensation. This is easy to overcome, as in the latter we considered only the simplest case, with $N = 2$ coherence domains in a cavity. Therefore, to account in a simple way for the correlations between the two water coherence domains, we can contemplate the two bosons coupled by a harmonic potential so as to form a "diatomic molecule" oscillating with pulsation $\omega = 2\pi\nu$ (where $\nu$ is the frequency) – more precisely, a three-dimensional isotropic oscillator (Gol'dman et al 1960, Landau and Lifshitz 1966, Messiah 1969). The two particle system may be formally described as a uniparticle system in a central field of force, but its basic physical nature essentially involves the two particles of masses $m_1$ and $m_2$ located at positions $r_1$ and $r_2$ and with their centre of mass immobile with respect to the cell. The problem is reformulated in the centre of mass coordinate system, supposed immobile with respect to the cell, and here we are concerned of the relative distance $d$ between the two coherent domains and of their reduced mass $\mu$:

$$d = r_1 - r_2 \quad \mu = \frac{m_1 m_2}{m_1 + m_2} = \frac{m_{eff}}{2} \tag{19}$$

Here $m_{eff} = m_1 = m_2$ is the 13.6 eV effective mass of each coherence domain; note in addition that $d = 2a$ when $r_1 = -r_2 = a$. The harmonic potential energy is of the form:

$$V(d) = \frac{1}{2}\mu\,\omega^2\,d^2 = \frac{1}{2}k\,d^2 \tag{20}$$

The probability distribution of the spherical harmonic oscillator in the position space is described generally by wavefunctions of the form (Gol'dman et al 1960):

$$\Psi_{n_r lm}(d) = d^l \cdot \exp\left[-\frac{d^2}{2(\hbar/\mu\omega)}\right] \cdot F\left(n_r, \; l + 3/2; \; \frac{d^2}{d_0^2}\right) \cdot Y_{lm}(\theta, \varphi) \tag{21}$$

where a characteristic distance $d_0$ is defined,

$$d_0 = \sqrt{\hbar/\mu\omega} = \sqrt{2} \cdot \sqrt{\hbar/m_{eff}\,\omega} \qquad (22)$$

$n_r$, $l$ and $m$ are quantum numbers,

$n_r = 0, 1, 2, ...$ with $n_r = n - (l + 1)$
$l = 0, 1, ... \, n_r = 0, 1, ... \, n - 1$
$m = -l, ... +l$

and F($-n_r$, $l + 3/2$; $d^2/d_0^2$) are the hypergeometric functions, while $Y_{lm}$ are the spherical harmonic functions. In the ground state, $n_r = 0$ ($n = 1$), $l = 0$ (no orbital movement), and $m = 0$, so that the corresponding wavefunction reduces to the simpler form:

$$\Psi_{000}(d) = \cdot \exp\left[-\frac{d^2}{2(\hbar/\mu\omega)}\right] \cdot F\left(0, \; 3/2; \; d^2\Big/d_0^2\right) \cdot Y_{00}(\theta, \varphi) \qquad (23)$$

which in fact is a Gaussian

$$\Psi_{000}(d) = \frac{1}{\sqrt{4\pi}} \cdot \exp\left[-\frac{d^2}{2(\hbar/\mu\omega)}\right] \qquad (24)$$

of halfwidth $d_0$, because F(0, 3/2; $d^2/d_0^2$) = 1 as we have $n_r = 0$, and $Y_{00}(\theta, \phi) = (4\pi)^{-1/2}$. Note also that the halfwidth $d_0$ is the incertitude or the standard deviation of the mean of the length $d$ measured between the two CDs in the ground state, $d_0 = \sigma_d = (<d>^2 - <d^2>)^{1/2}$ (while the corresponding position incertitude of each CD is $2^{-1/2}d_0$), and that $\Psi_{000}$ has the same form as the ground state of the onedimensional linear oscillator. It is obvious that, in order to accommodate the oscillator, a spherical cell - where the influence of the potential barrier represented by the plasmatic membrane will be neglected to a first approximation - must have a diameter *2a* proportional to the halfwidth $d_0$,

$$2a = cd_0 = c\sqrt{\hbar/\mu\omega} = c\sqrt{2} \cdot \sqrt{\hbar/m_{eff}\,\omega} \qquad (25)$$

where *c* is an adimensional constant ($c \geq 1$). In order to reduce as much as possible the perturbation due to the cell membrane, we take $c = 4$ covering thus a $4\sigma$ interval, so that the probability of finding the oscillator inside the cell will exceed 99.99 %. In this way a relationship between the cell radius *a* and the eigenfrequency $\omega$ was established,

$$a = 2\sqrt{2} \cdot \sqrt{\hbar/m_{eff}\,\omega} \qquad (26)$$

For instance, at an oscillation frequency $\nu$ of, say, $10^{11}$ s$^{-1}$, we may expect a cell radius of 18.7 $\mu$m – a value which is of the order magnitude of the eukaryotic cell dimensions.

Eq. (26) shows that the higher the frequency of the oscillations, the smaller the dimensions of the cell and, therefore, we may look after a criterion for a minimum frequency, allowing thus an estimate of a maximum cell radius. Assuming that in the living state the oscillator finds itself in the ground state of energy $E_{000} = 3\hbar\omega/2$ and that, in order to preserve the stability of the system, the thermal energy should be less than the quantum $\hbar\omega = E_{100} - E_{000}$ which rises the oscillator to the first excited state:

$$\hbar\omega \geq \frac{3}{2} k_B T \tag{27}$$

Then the *minimum frequency* is defined at a given temperature when the equality holds, $\omega_{min} = 3k_B T/2\hbar$. Thus for T = 310 K we get a minimum frequency $\nu_{min} = \omega_{min}/2\pi = 9.7 \times 10^{12}$ s$^{-1}$, which is contiguous to the Froehlich range, although slightly above its upper limit. Among other, this value shows that the previous frequency of $10^{11}$ s$^{-1}$ was too low, and implies also that the radius of 14.0 μm was too big. Indeed by substituting ω from eq. (27) into eq. (26) we get for the radius of the cell:

$$a \leq \frac{4}{\sqrt{3}} \cdot \frac{\hbar}{\sqrt{m_{eff} \, k_B T}} \tag{28}$$

and, with the numerical values for $m_{eff}$ and *T,* we find $a \leq 0.756$ μm for the cell radius. This value is about 3/4 of the maximum radius of 1.02 μm estimated in the spherical potential well model examined above. We admit that the two values compare rather well with each other, taking into account the approximations used, even though the corresponding maximum volume $V \leq 1.81$ μm$^3$ is significantly lower than the previous one of 4.45 μm$^3$ estimated for the spherical well. However, the predictions of the two models are of the same order of magnitude, which still is remarkable, as the harmonic oscillator is a two-particle system while the spherical well is a uniparticle one. Also, provided the relatively good agreement of the two values, most of the discussed biological inferences remain qualitatively unchanged. Nevertheless, the allowed cell volume interval 0.42 μm$^3 = V_{min} < V_{cell} < V_{max} = 1.81$ μm$^3$ narrowed substantially, which restricts the list of cell types fitting between these limits. This may suggest that the isotropic oscillator is a less adequate model for estimating the maximum cell volume as compared to the spherical well model. But the merit of the oscillator model lies perhaps not so much in its numerical predictions as in its two-particle nature, which allowed us to account for an interaction between the water coherent domains. In addition, the harmonic oscillator is a versatile model which might prove useful in a possible attempt to accommodate cells with various characteristics. The nature of the harmonic forces which couple the coherent domains has been neglected so far, but we may plausibly suppose that it is the same as the nonlinear and specific dielectrophoretic forces (eq. 1) proposed by Del Giudice et al (1983). For $\nu_{min} = 9.7 \times 10^{12}$ s$^{-1}$, the force constant $k = \mu\omega^2$ is very small, that is $3.8 \times 10^{10}$ times lower than in the HCl molecule. This uncommon ratio is however plausible, as the mass of the water coherence domain 'molecule' is about ~$10^9$ times lower as compared to the hydrochloric acid. Implicitly, the forces between the coherence domains seem to be considerable with respect to the mass of the former. Moreover, they are assumed to be long-range, Coulomb forces of the $r^{-3}$ form (Tuszynski 2004), a feature which is essential for giving material support to a large scale collective dynamics inside the cell.

*A synopsis on the simple quantum models of water coherent domains in the cell*

Our simple quantum models based on an effective mass of about 13.6 eV of the coherent polarization domains of water and on various, sometimes coarse, approximations succeeded to provide plausible interpretations of different relevant biological facts – the minimum and maximum volumes of prokaryotic cells and of some cellular organelles in the eukaryotes, as well as the harmful effects of $D_2O$ affecting mainly the eukaryotic cells. With respect to the cell dimensions, it is noteworthy that three different models, i.e. the Bose condensation of two coherence domains, the spherical potential well and the harmonic oscillator, yielded values of the same order of magnitude. Moreover, the specific energies involved in the translational and oscillatory movement of the CDs inside cells of ordinary dimensions fall in the Fröhlich frequencies band. The last two aspects, as well as the rather satisfactory match of the results to the biological parameters, entirely took advantage on the option of carrying out our calculations for quasiparticles with an effective mass as low as 13.6 eV, as predicted by the QED theory of the Milano group (Del Giudice et al 1986, Preparata 1995) on the basis of water molecule's physical properties. Moreover, as the above models connect the cellular data to the water coherence domains' effective mass, they provide new, though indirect, experimental support for the underlying CD theory, as coming from the field of cell biology and joining thus the large body of existing evidence (Blackman et al 1985, Bercu et al 2006, Fritz 1984, Grundler and Keilman 1983 and 1989; Liboff et al 1987; Liboff 1997, Novikov and Fesenko 2000, Paul et al 1983, Pohl 1983, Rowlands et al 1981, Rowlands 1983, Webb et al 1977, Webb 1980, Zhadin et al 1998). At the same time our models may benefit of their simplicity for suggesting the design of the necessary direct, physical and biophysical experimental proofs. Moreover we expect that much more biologically significant inferences still could be done by extracting the available physical information from the present simple quantum models and by refining and diversifying them, so as to account better for the huge variety of the living cell phenomenology.

It is noteworthy that the present order-of-magnitude or better agreement to the biological characteristics of certain cells was obtained notwithstanding the fact that in our models we assumed a spherical shape which does not fit the real shape for most cells or organelles, and that the specific internal structures as well as the molecular structure of membranes were neglected. This suggests that to a first approximation the dynamics of water coherence domains inside cells is weekly dependent on the organic molecules framework on which the system's architectonics is built and emphasizes thus the vital role of water in the living cell. In a biological perspective, we note finally that the outlined image is consistent to the current hypotheses of evolutionary internalization of certain organelles as small foreign cells, of metabolism pre-eminence over reproduction, and of the smallness and simplicity of the first protocells.

*Metastability vs. "immortality": a heuristic addendum*

As already mentioned, our simplified assumptions imply a treatment of water CDs as stable microscopic objects at thermal equilibrium, consistent to the picture of Del Giudice (1993). At the same time, they are objects consistent to the Fröhlich coherent electric oscillations, and therefore implying a nonequilibrium, metastable nature of the dissipative structures type (Prigogine 1980). Although in our models we do not account for dissipativity, we note that the latter has been shown to be the macroscopic manifestation of a microscopic invariance law (Del Giudice et al 1985). However, we may wonder if in the cell looked at as a simple quantum system, the microscopic stability of the water CDs, at its upper limit, would not entail time reversibility, and thus "immortality" (Dyson 2000)? This question is obviously beyond the scope of our approach, but we may allow for a brief discussion. Note that we supposed for the CDs in the living cell a role as some kind of Maxwell's demons; but Maxwell's demons should be metastable structures by virtue of the second law of thermodynamics (Wiener 1961); therefore assuming the stability of the water coherence domains implies that they should enter as components in larger metastable structures. In fact energy exchange processes between the water coherent domains and the Davydov solitons propagation along chain-like biopolymers (Davydov 1973, 1977, 1982) have been considered and investigated in detail (Del Giudice et al. 1982, 1985). Cytoskeleton filamentous proteins with ATPase activity controlled by water coherent domains were assumed to be involved in this interplay. However in time, the energy exchanges would be hampered by loss of enzymatic activity of the macromolecular chains associated to the slow but irreversible damage of chemical structure and spatial conformation (*i.e.*, ageing) of the former. Consequently, we may speculate that in the living cell it may happen that some of the CDs would be eventually annihilated (and probably, decay non-exponentially after a characteristic lifespan, i.e. not with a half-life time as in the common exponential decay) (*e.g.*, Gamow 1962). In such an event, the energy and impulse of the disintegrated CDs would be communicated to the surrounding incoherent phase of water, exciting vibrations of the molecules. On the other hand we suppose that the decay of CDs would lead to ageing and death of the cell; no cell is *stricto sensu* "immortal"; for example, in a "immortal" colony of bacteria the two daughter cells can be regarded as two new organisms or two "rejuvenated" copies of the parent cell because damaged macromolecules have been split between the two cells and diluted (Wikipedia 2008 - Biological immortality). The ensuing irreversibility and metastability of life would thus emerge as a consequence of the irreversible chemistry of proteins in water environment at ordinary temperature, largely in agreement to the image of molecular biology; however water appears to play a much more important role than admitted by the latter. These qualitative considerations argue for the necessity of further developments able to seize more accurately the cell dynamics in its complexity.

# Charge density oscillations in a new model of collective dynamics in water

In the following we present a ionic plasma model of water based on its partial dissociation, providing thus a new interpretation for some remarkable equilibrium and nonequilibrium properties of this unique liquid.

## *Physical prerequisites and overview*

Although liquid water has been studied extensively (Chaplin 2008), its highly unusual and diverse features (Pauling 1982) which are also required for the existence of life are not fully understood. The same applies to its various interactions as well as to its kinematic and dynamic correlations. To account for some of its characteristics from a standpoint of collective dynamics, we examined the possibility of plasmon-like excitations in liquid water within a model presented in detail elsewhere (Apostol and Preoteasa 2008). Here we will sketch it briefly.

It is widely agreed that the water molecule in the liquid state preserves to a large extent its integrity, especially the directionality of the $sp^3$-oxygen orbitals, though it may be affected substantially by hydrogen bonds (Pauling 1982). As such, water has a molecular electric moment, an intrinsic polarizability and performs hindered rotations (librations) which may affect its orientational polarizability.

We examine here another possible component of the dynamics of the liquid water, as resulting from the dissociation of its molecule. Water has two $H - O$ bonds which make an angle of about $109^o$ in accordance with the tetragonal symmetry of the four hybridized $sp^3$-oxygen orbitals. The "spherical" diameter of $H_2O$ molecule is approximately 2.75 Å and the inter-molecular spacing in liquid water under normal conditions is similar, a ~ 3 Å. This suggest that the $H_2O$ molecule in the liquid state, while preserving the directionality of the oxygen electronic orbitals, might be dissociated. Dissociation models which assume $OH^- - H^+$ or $OH^- - H_3O^+$ pairs are well known. This indicates a certain mobility of hydrogens (and oxygens).

Here, instead of considering the splitting of a few $H_2O$ molecules in ions with unit electric charges as in the current models, we examine the hypothesis that the dynamics of many molecules in liquid water has a component consisting of $O^{-2z}$ anions of mass $M = 16$ *amu* and density $n$ and $H^{+z}$ cations of mass $m = 1$ *amu* and density $2n$, where $z$ is a (small) reduced effective electronic charge (1 amu = $1.7 \times 10$-24 g). Given this small charge transfer, the hydrogen and oxygen atoms interact by long-range Coulomb potentials $\varphi_{OO}(r) = 4z^2e^2/r$, $\varphi_{HH}(r) = z^2e^2/r$ and $\varphi_{OH}(r) = - 2z^2e^2/r$, where $-e = - 4.8 \times 10^{-10}$ *esu* is the electron charge and $r$ denotes the distance between the ions. At the same time, the atoms interact by short-range repulsive (hard-core) potentials $\chi$. This combined interaction leads to a $H^{+z} - O^{-2z}$ two-species ionic stable plasma. As it is well known, a classical plasma with Coulomb interaction only is unstable; the ions have a classical dynamics due to their large mass. It is shown that in the limit $z \rightarrow 0$ water may exhibit an anomalous sound-like mode besides both the ordinary (hydrodynamic) one and the non-equilibrium sound-like excitations governed by short-range interactions.

*Plasma oscillations with two species of ions and the sound anomaly of water*

To describe the plasma oscillations with the two species of ions $O^{-2z}$ and $H^{+z}$ we consider the change in density associated with a displacement vector $v$ in the former and a displacement vector $u$ in the latter. We note that the Fourier transforms of the coulomb potentials are given by $\varphi_{OO}(q) = 4 \varphi(q)$, $\varphi_{HH}(q) = \varphi(q)$ and $\varphi_{OH}(q) = - 2 \varphi(q)$, where $\varphi(q) = 4\pi z^2e^2/q^2$ and $q$ is the wavenumber (momentum). The equations of motion obtained from the Lagrange function $L = T - U$ ($T$ and $U$ are the kinetic and potential energy terms, respectively) are:

$$m\ddot{u} + 2n q^2(\varphi + \chi)u - n q^2(2\varphi - \chi)v = 0$$
$$M\ddot{v} + n q^2(4\varphi + \chi)v - 2n q^2(2\varphi - \chi)u = 0$$

(29)

where $n = N/V$ is the density of water molecules, $\chi$ the Fourier transform of hard-core potential, admitted the same for both species of atoms, and $u$ and $v$ are displacement vectors.

The solutions of these equations show that in the long wavelength limit q $\rightarrow$ 0 two branches of eigenfrequencies appear, one corresponding to plasmonic oscillations and another to sound-like waves. The frequency of the plasma oscillations are given by: $\varphi$

$$\omega_p^2 = \frac{16\pi n z^2 e^2}{\mu}$$

(30)

and that of sound-like waves with velocity $v_s$ by:

$$\omega_s^2 = \frac{9n\chi}{M+2m} q^2 = v_s^2 q^2 \qquad (31)$$

In the above $\mu = 2mM/(2m + M)$ is the reduced mass. The plasma oscillations are associated with anti-phase oscillations of the relative coordinate ($2mu + Mv = 0$), while the sound waves are associated with in-phase oscillations of the center-of-mass coordinate ($u - v = 0$).

The sound-like excitations eigenfrequency branch provides a possible explanation for the so-called sound anomaly of water. According to eq. (31), the velocity $v_s$ of the sound-like waves with frequency $\omega_2 \sim \omega_s$ in the present two-componet fluid $H^{+z} - O^{-2z}$ is:

$$v_s = \sqrt{9\,n\,\chi/(M+2m)} \cong 3000 \ m/s \qquad (32)$$

and is distinct of the ordinary sound with velocity given in our case by

$$v_0 = 1/\sqrt{\kappa\,n(M+2m)} \cong 1500 \ m/s \qquad (33)$$

where $\kappa$ is the adiabatic compressibility. We see that both $v_o$ and $v_s$ exhibit a weak isotopic effect, while their ratio

$$v_s/v_0 = 3n\sqrt{\kappa\chi} \qquad (34)$$

does not. From this ratio we get the short-reange interaction,

$$\chi \cong 7eV \cdot 10^{-3} nm^3 \qquad (35)$$

In fact the sound waves are distinct from the ordinary, hydrodynamic, equilibrium sound with a temperature-dependent velocity, and represent a non-equilibrium elementary excitation whose speed does not depend on temperature (the sound anomaly of water). In order to distinguish them from the hydrodynamic sound we propose to call the sound-like excitations derived here *density "kinetic modes"* or *"densitons"*. In a collision-like regime, the hydrodynamic sound propagates, and the sound-like excitations do not. In the collision-less regime, the opposite occurs. Indeed, the phenomenon of two-sound anomaly in water is well documented (Teixeira et al 1985, Santucci et al 2006). Neutron, X-ray, Brillouin and ultraviolet light scattering on water revealed the existence of a hydrodynamic sound propagating with velocity vo ~ 1500 m/s for smaller wavevectors and an additional sound propagating with velocity ~3000 m/s for larger wavevectors. In addtion, though both sound velocities do exhibit an isotopic effect, their ratio does not. Therefore we assign this additional, faster sound to the sound-like excitations derived above.

*Polarization and dielectric function*

When an external electrical oscillating field of frequency $\omega$ arising from a potential $\phi(\mathbf{r})$ is applied to liquid water, the energies of the $O^{-2z}$ and $H^{+z}$ ions change by the additional amounts $U_O$ and $U_H$, and the previous equations of motion (29) do no more cancel but instead they become equal to $-izeq\phi$ and $2izeq\phi$, respectively, where $i$ is the imaginary unit. The water ionic plasma is therefore a system of coupled harmonic oscillators under the action of an external force and, in the limit of long wavelengths the solutions of the equations are given by:

$$u = \frac{i\,zeq}{m}\phi\,\frac{\omega^2 - \dfrac{2m}{3\mu}\omega_s^2}{\left(\omega^2 - \omega_p^2\right)\left(\omega^2 - \omega_s^2\right)}$$

$$v = -\frac{2i\,zeq}{M}\phi\,\frac{\omega^2 - \dfrac{2M}{3\mu}\omega_s^2}{\left(\omega^2 - \omega_p^2\right)\left(\omega^2 - \omega_s^2\right)}$$

(36)

The external field induces in water a polarization $P = - (E_u + E_v)/4\pi$, where the total internal field $E_u + E_v = E_{int}$ is related to the electrical induction $D = \varepsilon E = \varepsilon(D + E_{int})$ by the dielectric function (permittivity dispersion) $\varepsilon$. Here we disregarded the intrinsic and orientational polarizabilities. The internal fields are $E_u = -8\pi nzeu$ and $E_v = 8\pi nzev$ and we can calculate the dielectric function $\varepsilon$:

$$\varepsilon = 1 - \omega_p^2 / \omega^2$$

(37)

For frequencies lower than $\omega_p$ in eq. (37) the electromagnetic waves are absorbed (the refractive index is given by $n^2 = \varepsilon$). In fact water exhibits a strong absorbtion in the GHz – THz region (Tsai and Wu 2005, Padro and Marti 2003, Woods and Wiedemann 2004). On the other hand, neutron scattering on $D_2O$ (Bermejo et al 1995, Petrillo et al 2000) as well as inelastic X-ray scattering (Sette et al 1996) revealed the existence of a dispersionless mode $\sim 4-5$ meV ($\sim 10^{13}$ s$^{-1}$) in the structure factor, which may be taken tentatively as the $\omega_p$ plasmonic mode given by eq. (30). Making use of this equation we get

$$\omega_p \approx 3 \times 10^{14} z \ \ s^{-1}$$

(38)

for n = $1/a^3$, $a = 3$ A. Thus we may estimate the reduced effective charge $z \sim 3 \times 10^{-2}$; here we obtained an important numeric result as it gives a specific 'physical feeling' to our model.

An alternative estimate of z is possible from the static permittivity $\varepsilon_o$. To this purpose we modify eq. (37), where the exact cancellation of the external field by the internal field in the static limit ($\omega = 0$) was implicitly supposed to arise. More realistically, we assume that residual polarization fields due to intrinsic polarizability are still present in an external static field and give a contribution to the Debye model, which accounts mainly for an orientational polarizability of electric dipoles. In this way we took advantage of the compatibility of our ionic plasma model to the Debye model, which preserves also the directional character of $O - H$ bonds. Taking for the electric dipole $p = 2ez_e(a/2) = ez_ea$, where $a \sim 3$ Å and $z_e$ is a delocalized reduced charge associated with the $H - O$ dipole, together with $pD/T \ll 1$, we get for the static permittivity:

$$\varepsilon_0 = b = \frac{1}{1 - 4\pi\, n\, p^2 / 3T} \qquad (39)$$

which is the Kirkwood formula (Fröhlich 1958, Debye 1945). For the experimental value $\varepsilon_o =$ 80, we get the delocalized reduced charge associated with the $H$-$O$ dipole, $z_e \sim 10^{-2}$. This value is in good agreement with $H^{+z} - O^{-2z}$ plasma charge $z$ evaluated above.

*Quantization, Debye screening and the correlation energy*

The plasma oscillations may be quantized and may contribute to the elementary excitations introduced recently in a model for the local, collective vibrations of particles in liquids with a two-dimensional boson statistics (Apostol 2006, 2007). The energy levels of the elementary excitations are:

$$E_n = \sum_q \hbar\, \omega_p\left(n+1/2\right) = \frac{V}{\left(2\pi\right)^3} \cdot \frac{4\pi}{3}\, q_c^3 \cdot \hbar\, \omega_p\left(n+1/2\right) \qquad (40)$$

where $q_c$ is a cutoff wavevector which we take as $q_c \sim 1/a$, and $\omega_p$ is the plasma oscillation frequency given by eq. (30):

$$\omega_p \approx 200\, z\ meV \qquad (41)$$

The plasma excitations described above represent collective oscillations of the density in the long wavelength limit, which induce correlations in the ionic movements. As the Coulomb energy is much lower than the temperature of liquid water, $z^2 e^2/a \sim 45$ K for $z \sim 3 \times 10^{-2}$, for a classical plasma these correlations are associated with the Debye-Huckel screening length (Landau and Lifshitz 1980). This allowed an order-of-magnitude estimate of the correlation energy (vaporization heat) of water per particle:

$$\varepsilon_{corr} = -\frac{e^2}{a} \sqrt{\frac{\pi\, e^2}{T\, a}} \left(6\, z^2\right)^{3/2} \qquad (42)$$

giving $\varepsilon_{corr} \sim 10^2$ K at room temperature.

*The water ionic plasma in the general case*

Similar results are obtained for $OH^- - H_3O^+$ or $OH^- - H_3O^+$ dissociation modes of water, therefore the plasma model treated here may be looked at as an effective model for various ionic plasma components that may exist in liquid water.

The model is generalized for asymetric short-range interactions, by considering a $H^{+z} - O^{-2z}$ plasma where the short range interactions $\chi_{ij}$ are different for the $OO$, $OH$, $HH$ pairs. For a non-separable short-range interaction, the spectrum may exhibit multiple branches and may serve to identify the mass and charge of various molecular aggregates in a multi-component plasma.

*Biological implications*

In particular, a ionic plasma of water may occur *e.g.* in the cell cytoplasm, where many types of ionic groups exist. Moreover the calculated spectrum may be correlated with various data which evidenced specific effects in living cells in the $10^{10} - 10^{12}$ Hz frequency range by microwave, Raman and optical spectroscopy and by cell biology studies (Webb 1980, Webb et al 1977, Rowlands 1981, Bercu et al 2006), and which sustain the theory of coherent domains in living matter (Fröhlich 1968b, 1975, 1978, 1980, 1982, Del Giudice et al 1982, 1986, 1988, 2002). It may also be relevant with respect to the so-called Zhadin effect (Blackman et al 1985, Liboff et al 1987; Liboff 1997), *i.e.* sharp conductivity changes in amino acids solutions (Zhadin et al 1998) and protein hydrolysis when exposed to ion cyclotron resonance (ICR) magnetic fields tuned to $q/m$ of the amino acids (Novikov and Fesenko, 2000), as well as the pH changes of water in electric fields applied from outside (Del Giudice 2008). In the cell, the ionic plasma oscillations of water and the fields they generate may interact with various electric fields associated to biomembranes, biopolymers and water polarization coherence domains and, as such, they may play a certain role in intra- and intercellular communications. We note that the water ionic plasmons should have a very low excitation energy (effective mass), of ~$200 \cdot z$ meV, and they are almost dispersionless. Accordingly, the associated de Broglie wavelength may be very large, and entanglement of their wavefunctions may be possible. This might provide support for intercellular correlations at very long distance, of major interest for biological phenomena such as embrio-, angio-, and morphogenesis, malign proliferation, contact inhibition, tissue repair, etc. Finally we note that a model of physical plasma has been proposed recently (Zon 2005) for the brain, which is considered to be a seat of vast number of subvolumes containing mobile charge carriers, able to undergo phase transitions between states of noncollective and collective behavior.

# Conclusions

Based on the concept of collective dynamics in water, the models presented above were able to offer explanations, or at least plausible interpretations, for phenomena as different as the dimensions of cells, the toxicity of heavy water and the physical properties of liquid water. Both the coherent domains of water advanced by the QED theory of coherence in condensed matter and the sound-like densitons proposed here for the ionic plasma oscillations are elementary excitations built on long-range electromagnetic interactions and showing specific frequencies in the Fröhlich band.

Taking advantage of the wavelike properties of water coherent domains due to their low effective mass, we showed that simple quantum models of the cell may explain qualitatively as a minimum some cellular characteristics or even give numerical results in satisfactory agreement to biological data. The boson gas condensation hypothesis predicted a minimum cell volume of about 0.4 $\mu m^3$, while for the maximum cell volume values of 4.5 and 1.8 $\mu m^3$ were estimated by the models of spherical well with impenetrable walls and of isotropic oscillator, respectively. Within the individual variability, cellular volumes of $1 - 5$ $\mu m^3$ of certain typical prokaryotic cells such as *Eubacteria* and *Myxobacteria* and some smaller species of *Cyanobacteria*, as well as volumes of $0.7 - 8$ $\mu m^3$ of mitochondria and of cisternae of the Golgi apparatus in the eukaryotic cells give straight support for the range of volumes defined by our models, in spite of their coarse approximations. A maximum cellular volume of 0.18 $\mu m^3$ evaluated for a particular case of spherical well with semipenetrable walls fits well to the size of the smallest prokaryotes, e.g. *Mycoplasma* and *Rickettsia* (0.01 − 0.03 $\mu m^3$), the blue-green alga *Prochlorococcus* (0.1 − 0.3 $\mu m^3$), etc., although in this case the minimum volume estimate by the boson gas condensation model does not apply, evidencing possible limits of our present picture for the dynamics of water in the smallest cells. However the spherical well with semipenetrable walls describing the water coherence domains inside the compartmentalized eukaryotic cells is able to explain the toxic effects of heavy water affecting these cells more than the prokaryotes.

In the model of liquid water based on plasmon-like excitations, long-range (Coulomb) and short-range potentials yield a $H^{+z} - O^{-2z}$ ionic stable plasma, characterized by two branches of eigenfrequencies, one corresponding to plasmonic oscillations and another to sound-like waves. The sound waves are distinctive from the ordinary, hydrodynamic sound, and represent a non-equilibrium elementary excitation whose speed does not depend on temperature - the sound anomaly of water known from neutron, X-ray, Brillouin and ultraviolet light scattering on water. In an external electrical oscillating field, the dielectric function (permittivity dispersion) showed lower frequency absorption, as evidenced by water in the GHz-THz region. Taking the dispersionless mode of $\sim 10^{13}$ $s^{-1}$ (4 − 5 meV) revealed by neutron and inelastic X-ray scattering as the plasmonic mode, we obtained $z \sim 0.03$, in agreement to the delocalized reduced charge of the *H-O* dipole, $z_e \sim 0.01$, deduced from the static permittivity. The plasma oscillations may be quantized, allowing an estimate of the correlation energy (vaporization heat) of water. The results are generalized for a multi-component plasma, as in the cell cytoplasm where many types of ionic groups exist. In the living cell, the ionic plasma oscillations may interact with fields of inner structures and water coherence domains. Our results may be correlated to microwave effects, to Raman and optical spectroscopy of living cells, to biological data, and to the more recent ionic cyclotron resonance experiments. As the water ionic palsmons should have very low effective mass and very large de Broglie wavelength, their wavefunctions may undergo entanglement, supporting thus intercellular correlations at very long distance in macroscopic organisms.

Both the simple quantum models of coherence domains in the cell and the plasmon model of liquid water allow substantial further developments and thus accommodate more cell biology facts. Considering their possible role in the living cell, they are consistent to the Fröhlich picture which transcends the limits of current molecular biology, and they may explain cell phenomena not fully understood by the latter.

Nevertheless, our approach is not contradictory but complementary to molecular biology. It totally fits the image proposed by Del Giudice (1993) for the living state, which „charts its own way through a close interplay between chemical and electromagnetic coherence" and which can be described as a superposition of a „coherent phase" whereby the particles loose their individual identity, cannot be separated, move together with each other, and are kept in phase by a trapped EM field, and of an „incoherent phase" in which the particles are localizable and separable, each moves in spite of the others, and are bound together by static, short-range forces at most.

Therefore, our results sustain also that in addition to the analytical investigation of cells at molecular level as currently done in molecular biology, understanding life requires physical methods able to interfere with the EM fields involved in the collective dynamics of the composing molecules. They fully support "the electromagnetic paradigm of living organism" proposed by Liboff (2004, 2005) in search for a rational biology and medicine.

## Acknowledgement

We would like to pay tribute to the memory of George Emil Palade, who made so much in the discovery of cellular ultrastructure but who imparted the belief that a real insight of the living cell is to be searched beyond its molecular architectonics. One of us (E.A.P.) gratefully thanks Professors Emilio Del Giudice (Milano) and Abraham R. Liboff (Boca Raton) for valuable discussions, fascinating ideas and generous encouragement, as well as Prof. Jiri Pokorny (Prague) for warm appreciation. Many thanks are due to Dr. Vladimir Gheordunescu (Bucharest) for relevant biological data and helpful discussions. Contract / grant sponsor: Romanian Ministry of Education and Research.